# Scaling-up atomically thin coplanar semiconductor-metal circuitry via phase engineered chemical assembly

Xiaolong Xu[1,2], Shuai Liu[1], Bo Han[3], Yimo Han[4], Wanjin Xu[1], Xiaohan Yao[1], Kai Yuan[1], Pan Li[1], Shiqi Yang[5], Wenting Gong[1], David A. Muller[4], Peng Gao[2,3], Yu Ye[1,2,6]* and Lun Dai[1,2,6]*

[1]State Key Laboratory for Artificial Microstructure &Mesoscopic Physics, School of Physics, Peking University, Beijing 100871, China

[2]Collaborative Innovation Center of Quantum Matter, Beijing 100871, China

[3]Electron Microscopy Laboratory, School of Physics, Peking University, Beijing 100871, China

[4]Kavli Institute at Cornell for Nanoscale Science, Cornell University, Ithaca, New York 14850, USA

[5]Academy for Advanced Interdisciplinary Studies, Peking University, Beijing 100871, China

[6]Nano-optoelectronics Frontier Center of Ministry of Education, Peking University, Beijing 100871, China

*Correspondence and requests for materials should be addressed to Y.Y. (email: ye_yu@pku.edu.cn) and L.D. (email: lundai@pku.edu.cn)

**Two-dimensional (2D) layered semiconductors, with their ultimate atomic thickness, have shown promise to scale down transistors for modern integrated circuitry. However, the electrical contacts that connect these materials with external bulky metals are usually unsatisfactory, which limits the transistor performance. Recently, contacting 2D semiconductors using coplanar 2D conductors has shown promise in reducing the problematic high resistance contacts. However, many of these methods are not ideal for scaled production. Here, we report on the large-scale, spatially controlled chemical assembly of the integrated 2H-MoTe$_2$ field-effect transistors (FETs) with coplanar metallic 1T′-MoTe$_2$ contacts via phase engineered approaches. We demonstrate that the heterophase FETs exhibit ohmic contact behavior with low contact resistance, resulting from the coplanar seamless contact between 2H and 1T′ MoTe$_2$**

**confirmed by transmission electron microscopy characterizations. The average mobility of the heterophase FETs was measured to be as high as 23 cm$^2$ V$^{-1}$ s$^{-1}$ (comparable with those of exfoliated single crystals), due to the large 2H MoTe$_2$ single-crystalline domain (486±187 μm). By developing a patterned growth method, we realize the 1T′ MoTe$_2$ gated heterophase FET array whose components of channel, gate, and contacts are all 2D materials. Finally, we transfer the heterophase device array onto a flexible substrate and demonstrate the near-infrared photoresponse with high photoresponsivity (~1.02 A/W). Our study provides a basis for the large-scale application of phase-engineered coplanar MoTe$_2$ semiconductors-meter structure in advanced electronics and optoelectronics.**

Integrated two-dimensional (2D) electronic circuits based on 2D layered semiconductors, including the transition-metal dichalcogenides (TMDCs) as well as other 2D semiconductors, such as atomically thin black phosphorus, InSe, and Te etc., keep the promise for advanced electronics and flexible devices with increased functionality, performance, and scaling in integrated circuits[1-7]. The electronic and optoelectronic devices performance is significantly affected by the characteristics of the electrical contacts that connect the semiconductor materials with external circuitry[8-9]. As the pristine surface of a 2D material has no dangling bonds, it is difficult to form strong interface bonds with a metal, thereby increasing the contact resistance[8]. Moreover, direct metal electrodes deposition on the 2D semiconductor surface causes considerable defects, strain, disorder, and metal diffusion, resulting in a glassy layer dominated by inter-diffusion and strain[10]. Previous efforts, e.g. aligning the metal work function with the conduction/valence band edge of 2D layered semiconductors, showed unsatisfactory high-resistance contacts, due to Fermi level pinning[8]. The traditional method to reduce the contact resistance for silicon is to decrease the depletion region width by locally doping near the silicon-metal junction. However, the 3D doping method used in silicon technology cannot be employed in 2D devices. Recently, contacting 2D semiconductors using coplanar 2D conductors has shown promise in reducing the problematic high resistance contacts. Reduced contact resistance were observed in metallic 1T MoS$_2$/semiconducting 2H MoS$_2$, metallic 1T′ MoTe$_2$/semiconducting 2H MoTe$_2$, and metallic VS$_2$/semiconducting 2H MoS$_2$ with seamless coplanar interfaces, which were fabricated using intercalation, laser heating induced phase transition, and heteroepitaxy growth, etc[11-14]. However, many of these methods are not ideal for scaled production. Thus, spatially controlled integration of metallic and semiconducting 2D materials in a large scale is of significant importance for practical applications. Previous studies revealed that $MoTe_2$ is particularly interesting for phase-engineering applications, because the free energy difference between semiconducting 2H $MoTe_2$ and metallic 1T′ $MoTe_2$ is much smaller (~35 meV per $MoTe_2$ formula unit) compared with those of other TMDC materials[18-21]. This small energy difference leads to the possibility of phase controlled synthesis in a large scale.

Here, we demonstrate the large-scale, spatially controlled chemical assembly of the integrated 2H $MoTe_2$ field-effect transistors (FETs) with coplanar metallic 1T′ $MoTe_2$ contacts via phase engineered approaches. The heterophase FETs exhibits ohmic contact behavior with a low contact resistance of ~1.1 kΩ μm at high doping level and high average carrier mobility of ~23 $cm^2 V^{-1} s^{-1}$. We also developed a patterned growth method of 1T′ $MoTe_2$ and realized the 1T′ $MoTe_2$ gated heterophase FET array whose components of channel, gate and contacts are all 2D materials. Each FET has an independent patterned 1T′ $MoTe_2$ gate electrode, which is required for logic circuitry. Few-layer 2H $MoTe_2$, with a band gap of ~1 eV, is a candidate material for near-infrared (NIR) photodetecting[15-17]. We transfer the heterophase device onto a flexible substrate, and demonstrate the NIR photoresponse with enhanced photoresponsivity (~1.02 A/W) due to the reduced contact resistance. Our study promises the potential large-scale applications of 2D coplanar heterophase structure in advanced electronic and optoelectronic devices.

**Results**

**Patterned chemical assembly of the heterophase $MoTe_2$ in a large sclae**

In order to chemically assemble large-scale coplanar 2H-1T′ heterophase $MoTe_2$, high-quality few-layer 2H $MoTe_2$ film was first synthesized on a $p^+$-Si/$SiO_2$ substrate via the chemical vapor deposition (CVD) method (Fig. 1a). Our previous work has found the synthesis of 2H $MoTe_2$ film was driven by the solid-to-solid 1T′ to 2H $MoTe_2$ phase transformation[22], which can be well described by the time-temperature-transformation diagram. By controlling the kinetic rates of nucleation and crystal growth, we were able to synthesize large-scale continuous 1T′ $MoTe_2$ and 2H $MoTe_2$ thin film, as well as arbitrary shaped 1T′ $MoTe_2$ and 2H $MoTe_2$

(see Supplementary Section Ia). Considering the time consuming, we synthesized a centimeter-scale 2H-MoTe$_2$ thin film with a domain size up to several hundred micrometers under 650 °C for two hours (Fig. 1e) (see details in Methods). The large-scale electron backscatter diffraction (EBSD) map of the as-synthesized 2H MoTe$_2$ thin film shows a uniform contrast in the out-of-plane direction (see Supplementary Section Ib), indicating that the highly textured MoTe$_2$ thin film stacked in the c-axis, with an average domain size up to 486±187 μm in the in-plane direction (Fig. 1i). The domain size ($d$) is calculated based on a circularly equivalent area ($d = \sqrt{4s/\pi}$), where $s$ is the statistically averaged domain area. Then, large-scale 2H MoTe$_2$/Mo striped period was fabricated by photolithography, reactive ion etching (RIE), and magnetron sputtering Mo, followed by a lift-off process (Fig. 1b and 1f). Subsequently, the sample was sent back into the furnace at lower temperature (530 °C) for 30 min. After the second CVD growth, the Mo film was tellurized to 1T′ MoTe$_2$ while the 2H MoTe$_2$ remains unaffected (see Supplementary Section Ic). Therefore, the 2H/1T′ MoTe$_2$ striped patterns were obtained (Fig. 1c,g). This is confirmed by the appearance of the well-resolved A$_g$ Raman modes (107 cm$^{-1}$, 127 cm$^{-1}$, 161 cm$^{-1}$, and 256 cm$^{-1}$) of 1T′-MoTe$_2$ and the out-of-plane A$_{1g}$ (171 cm$^{-1}$), B$_{2g}^1$ (291 cm$^{-1}$), and the strong in-plane E$_{2g}$ (234 cm$^{-1}$) Raman modes of 2H MoTe$_2$ in the respective regions (Fig. 1j). This heterophase pattern was further fabricated into isolated 1T′-2H-1T′ MoTe$_2$ arrays by photolithography, and reactive ion etching (RIE, Fig. 1d and 1h). Using the electron beam lithography (EBL) in stand of the photolithography, we also synthesized arbitrary coplanar 2H/1T′ MoTe$_2$ heterophase structures with smaller featured size, e.g. Peking University (PKU) logo (Fig. 1k). Raman spectroscopy mapping of the representative fingerprints of the 2H MoTe$_2$ E$_{2g}$ mode and 1T′ MoTe$_2$ A$_g$ mode (labelled by the corresponding dashed lines in Fig.1j) clearly shows that both the 2H-MoTe$_2$ PKU logo and 1T′-MoTe$_2$ background are uniform, and the interfaces between them are sharp and clear (Fig. 1l).

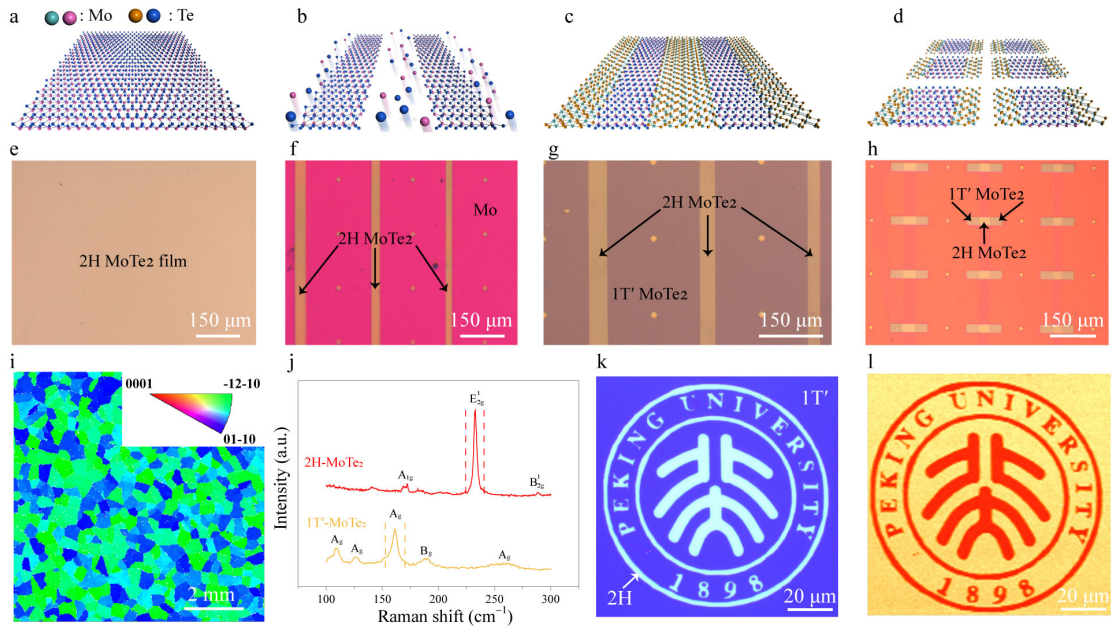

**Figure 1| Fabrication processes and Raman characterizations of the 2H/1T′ heterophase MoTe₂ coplanar structure. a-h,** Schematic illustration and corresponding optical images of the chemical assembly processes of 2H/1T′ heterophase MoTe₂. **(a, e)** High-quality large-scale 2H MoTe₂ film was first synthesized on $p^+$-Si/SiO₂ substrate via the CVD method. **(b, f)** 2H MoTe₂/Mo striped patterns were fabricated by photolithography, RIE, and magnetron sputtering Mo, followed by a lift-off process. **(c, g)** 2H/1T′ MoTe₂ striped pattern was obtained by a second CVD growth at a lower temperature (530 °C) for 30 min. **(d, h)** This heterophase structure was patterned into 1T′-2H-1T′ MoTe₂ array by photolithography and RIE process. **i,** A large-scale EBSD map of the as-synthesized 2H MoTe₂ thin film, indicating an average single-crystalline domain size up to 486±187 μm in the in-plane direction. **j,** Raman spectra of the heterophase MoTe₂ structure. The Raman spectrum of 2H MoTe₂ remains unchanged after the second CVD growth. The well-resolved Raman signatures of 1T′-MoTe₂ confirm that the Mo thin film was fully tellurized into 1T′-MoTe₂. **k,** Optical image of the PKU logo of 2H/1T′ MoTe₂ heterophase structure. **l,** Raman spectroscopy mapping of $E_{2g}^1$ mode of 2H MoTe₂ and $A_g$ mode of 1T′ MoTe₂. Both the 2H MoTe₂ PKU logo and 1T′ MoTe₂ background are uniform and the interfaces between them are sharp and clear.

**Characterizations of the interface of the coplanar 2H-1T′ heterophase MoTe₂**

The interface between 2H MoTe₂ and 1T′ MoTe₂ were characterized using atomic force microscopy (AFM) and high-resolution transmission electron microscopy

(HR-TEM). The typical optical image of an isolated 1T′-2H-1T′ MoTe$_2$ coplanar heterostructure shows a large contrast difference between respective regions (Fig. 2a), resulting from the clear dielectric constant difference between the semiconducting 2H MoTe$_2$ and metallic 1T′ MoTe$_2$[23]. The AFM height image at the 2H/1T′ MoTe$_2$ interface (Fig. 2b) exhibits a nearly homogeneous color contrast, indicating the uniformity of the sample thickness. The line profiles of the 2H MoTe$_2$ and 1T′ MoTe$_2$ (Fig. 2c) regions show a similar thickness of about 5 nm (corresponding to 7-layer MoTe$_2$[24]), resulting from the precise control of the Mo film thickness. Additionally, the line profile across the interface (see Supplementary Section II) does not show any stitches or flake overlaps, indicating the seamless coplanar contact between the heterophases. To evaluate the crystallinity of the heterophases, we transferred 1T′-2H-1T′ MoTe$_2$ arrays onto a copper grid (see Supplementary Section IIIa). At the interface of 2H/1T′ MoTe$_2$ (Fig. 2d), the selected-area electron diffraction (SAED) pattern (Fig. 2e) comprises a single set of diffraction spots with six-fold symmetry related to the single-crystalline 2H MoTe$_2$ together with a series of diffraction rings related to the polycrystalline 1T′ MoTe$_2$[25]. The HR-TEM image (Fig. 2f) shows the interface contains polycrystalline monoclinic (1T′) structure domains of about tens of nanometers and a single hexagonal (2H) structure domain. We further used high-angle annular dark-filed scanning TEM (HAADF-STEM) to resolve in-plane atomic arrangements of a typical 1T′-2H MoTe$_2$ interface (Fig. 2g), showing atomically smooth interface transition. The corresponding FFT pattern contains a set of monoclinic 1T′ spots and a set of hexagonal 2H spots (Fig. 2h). The simulated HAADF-STEM image as well as the schematic lattices (Fig. 2g) at the interface are provided, which agree well with the observed results. We found the 1T′ domains stitch seamlessly to single crystalline 2H MoTe$_2$ with random orientations at different location of the interface (see Supplementary Section IIIb).

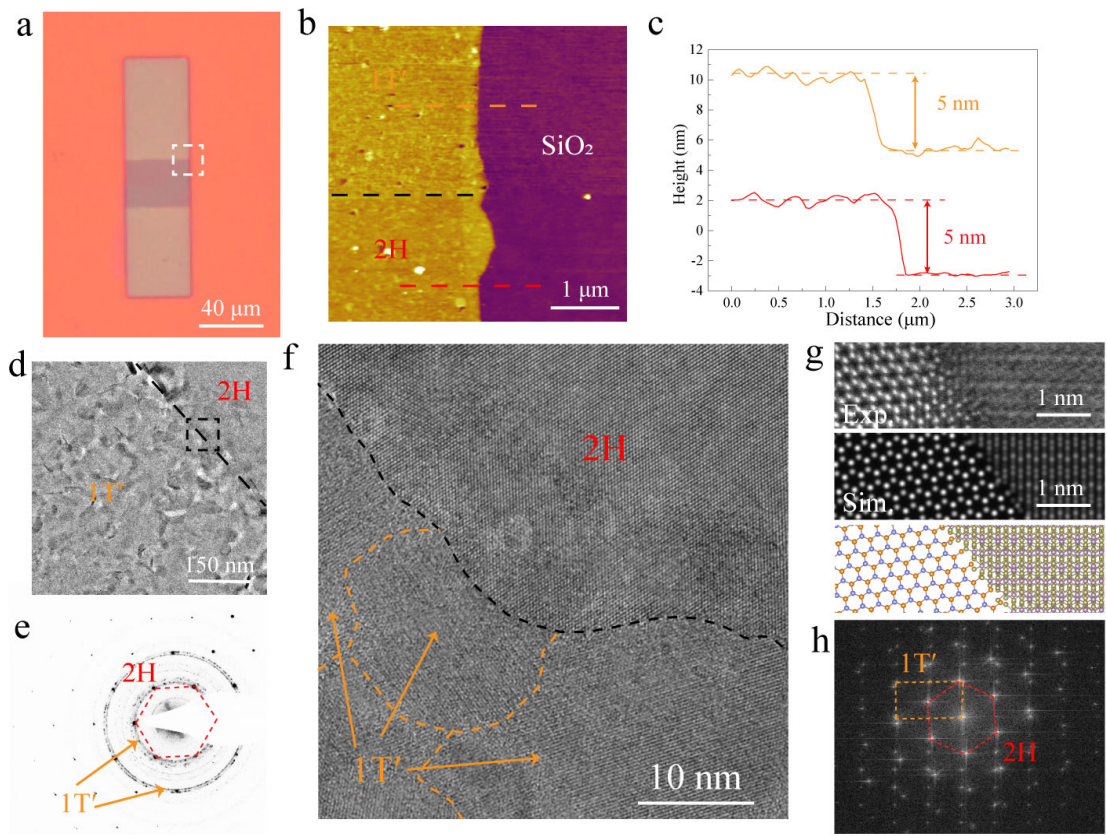

**Figure 2| AFM and HR-TEM characterizations of the 2H/1T′ MoTe₂ interface. a,** Optical image of a 1T′-2H-1T′ heterophase MoTe₂ structure. The interface area (labelled by the dashed box) was measured by the AFM. **b,** The AFM height image at the 2H/1T′ MoTe₂ interface. The nearly homogeneous color contrast indicates the uniform and identical sample thickness at the 2H and 1T′ MoTe₂ regions. **c,** The height line profiles at 2H MoTe₂ and 1T′ MoTe₂ regions, respectively, showing a same thickness of about 5 nm. **d,** Low-magnification TEM image at the 2H/1T′ MoTe₂ interface. **e,** SAED pattern at the interface, which comprises a single set of diffraction spots with six-fold symmetry related to the single-crystalline 2H MoTe₂ and a series of diffraction rings related to the polycrystalline 1T′ MoTe₂. **f,** HR-TEM image at the interface. 1T′ MoTe₂ domains with size of about tens of nanometer stitch seamlessly to 2H MoTe₂ with random orientations. **g,** Experimental and simulated HAADF-STEM images as well as the schematic lattices of a representative 1T′/2H MoTe₂ interface, showing atomically smooth interface and seamless contact. **h,** The corresponding FFT pattern of the experimental HAADF-STEM image, showing a set of monoclinic 1T′ diffraction spots and a set of hexagonal 2H diffraction spots.

**Electrical properties of the coplanar heterophase MoTe₂ FETs**

The seamless contact allows us to inject current into semiconducting 2H-MoTe$_2$ from metallic 1T′-MoTe$_2$ electrodes with reduced contact resistance ($R_c$). Besides, the large-scale chemical assembly of coplanar semiconducting and metallic MoTe$_2$ structure developed in this work is of great significance for practical applications. The large-scale coplanar heterophase MoTe$_2$ FETs array with different channel length were fabricated (Fig. 3a). Pd/Au electrodes (10/50 nm) were deposited on the 1T′ parts of each 1T′-2H-1T′ MoTe$_2$ structure for electrical measurement. The $p^+$-Si was used as the back gate electrode. The devices array with Pd/Au electrodes directly deposited on the 2H phase were fabricated for comparison (see Supplementary Section IVa). With the same channel length, the current injected from coplanar 1T′-MoTe$_2$ contact is one order of magnitude larger than that from the deposited Pd/Au metal contact (Fig. 3b), indicating lower contact resistance is formed at the 1T′/2H MoTe$_2$ interface. The typical source-drain current ($I_{ds}$) versus source-drain voltage ($V_{ds}$) curves measured at various gate voltages (Fig. 3c) show linear behavior, confirming the ohmic contact between the 1T′ and 2H MoTe$_2$. The gate voltage ($V_g$) dependence of the heterophase FET under bias voltage of 0.5 V shows the $p$-type channel characteristic with an on-off ratio of ~1×10$^4$ (Fig. 3d). The room-temperature field-effect mobility ($\mu$) is obtained to be ~32 cm$^2$ V$^{-1}$ s$^{-1}$, using $\mu = (dI_{ds}/dV_g)(L/W)(1/V_{ds}C_g)$, where $L$, $W$, and $C_g$ are the channel length, channel width, and the gate capacitance per unit area, respectively. We have measured 100 FETs with different change lengths, and found that the average mobility value (in a range of 20-24 cm$^2$ V$^{-1}$ s$^{-1}$, comparable to the reported values for exfoliated 2H MoTe$_2$ single crystals[12-13]) was independent of the channel length (Fig. 3e). This was not surprising, because in our experiment, the average single-crystalline domain size of the 2H MoTe$_2$ (~486 μm) was much larger than the largest channel length (60 μm) being measured.

To determine the contact resistance, $R_c$, between the 1T′ MoTe$_2$ and 2H MoTe$_2$, we used the transfer length method (TLM)[9], in which the channel length dependent resistances of both the 1T′-2H-1T′ coplanar heterophase structure (see Supplementary Secition IVb) and 1T′ MoTe$_2$ (see Supplementary Section IVc) were measured. The total $R_c$ of the metal-1T′-2H MoTe$_2$ structure contains the $R_c$ of metal-1T′ MoTe$_2$ and the $R_c$ of 1T′-2H MoTe$_2$. Because of the semiconducting nature of 2H MoTe$_2$, the total $R_c$ (metal-1T′-2H MoTe$_2$) shows strong dependence on the back gate. In contrast, due to the metallic nature of 1T′ MoTe$_2$, the $R_c$ of metal-1T′ MoTe$_2$ is ~0.6 kΩ μm, independent of the back gate (Fig. 3f). The minimum total $R_c$ is measured to be ~1.7 kΩ μm at a higher turn-on gate voltage (Fig. 3f). The corresponding $R_c$ for the 1T′-2H MoTe$_2$ is ~1.1 kΩ μm, which is about two orders of magnitude smaller than those of metal-contacted 2H-MoTe$_2$ (reported value of 409 kΩ μm)[13].

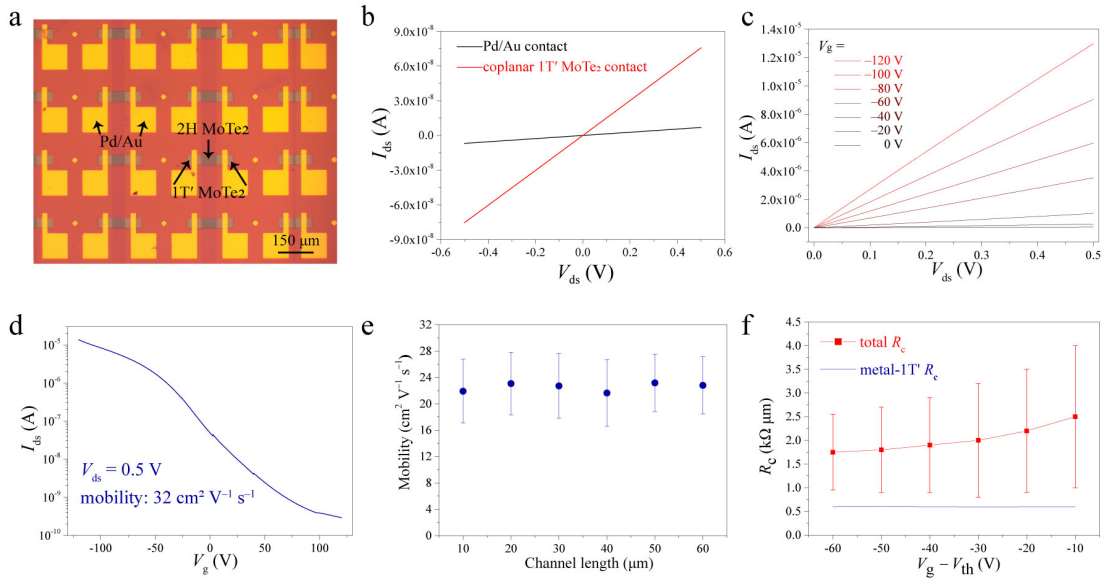

**Figure 3| Room-temperature electrical properties of the 2H/1T′ MoTe$_2$ coplanar heterophase FETs. a,** Optical image of the heterophase FET device array with different channel lengths. **b,** $I_{ds}−V_{ds}$ curves of a representative coplanar 1T′ MoTe$_2$ contacted and direct Pd/Au metal contacted 2H MoTe$_2$ channel under zero back gate voltage. The coplanar 1T′ MoTe$_2$ contact shows superior contact property. **c,** Typical $I_{ds}$-$V_{ds}$ curves of the heterophase FET measured under various gate voltages. All the $I_{ds}$-$V_{ds}$ curves exhibit linear ohmic contact behavior. **d,** Corresponding transfer curve of the heterophase FET, showing *p*-type channel characteristic with an on-off ratio of ~1×10$^4$. The room-temperature field-effect mobility is estimated to be ~32 cm$^2$ V$^{−1}$ s$^{−1}$. **e,** Field-effect mobilities for different channel length measured on 100 devices. All the mobilities hover in a range of 20-24 cm$^2$ V$^{−1}$ s$^{−1}$ and are independent of the channel length, due to the large single-crystalline domain of the 2H MoTe$_2$. **f,** Contact resistance of the heterophase and 1T′ MoTe$_2$ only devices extracted using the transfer length method. The total contact resistance shows strong gate dependence, while the contact resistance of the 1T′ MoTe$_2$ is about 0.6 kΩ μm, independent of the gate voltage. The measured minimum value of the total contact resistance is about ~1.7 kΩ μm, which is obtained at high doping level, yielding a coplanar 1T′-2H MoTe$_2$ contact resistance of ~1.1 kΩ μm.

**1T′ MoTe$_2$ gated heterophase transistor and NIR flexible photodetector**

In order to realize 2D circuitry, it is required to control each transistor independently. Herein, we chemically assembled the coplanar 1T′-2H-1T′ MoTe$_2$ array on top of the pre-patterned 1T′ MoTe$_2$ gate electrodes (Fig. 4a). The 1T′ MoTe$_2$ patterns were synthesized via the developed patterned growth method (see Supplementary Section Ia). A layer of 30 nm Al$_2$O$_3$ formed by atomic layer deposition (ALD) was used as the gate dielectric. The detailed fabrication process is provided in the Method and Supplementary Section IVd. For a typical 1T′ MoTe$_2$ gated heterophase transistor (inset of Fig. 4b), the *I-V* curves measured at various gate voltages show linear behavior (Fig. 4b), indicating the ohmic contact between the coplanar 1T′-2H MoTe$_2$. The transfer curve exhibits a clear *p*-type channel characteristic (Fig. 4c). Herein, the on-current of the transistor is about one order of magnitude lower compared with the device depicted in Fig. 3a. The reduction of the on-current may result from the *n*-type doping effect induced by the ALD Al$_2$O$_3$ (see Supplementary Section IVe)[26]. It is reasonable to envision that the device performance can be further improved by replacing Al$_2$O$_3$ layer with *h*BN[27-28].

With a bandgap of ~1.0 eV, few-layer 2H MoTe$_2$ keeps the promise for the near-infrared (NIR) photodetection[17]. The chemically assembled coplanar 1T′-2H-1T′ MoTe$_2$ heterophases with reduced contact resistances might result in improved photodetection performance in a large scale. Herein, we transferred the large-scale 1T′-2H-1T′ MoTe$_2$ array to a flexible polyimide (PI) substrate (Fig. 4d, see Method for details). The heterostructure maintains its integrity in large scale after the transfer (Fig. 4e). We investigated the photoresponse behavior of the heterophase under 1064 nm laser illumination with various incident light powers. The photocurrent increases with the incident light power (Fig. 4f). It is worth noting that when the incident 1064 nm laser power is 50 nW, the photoresponsivity ($R = \Delta I/P$, where $\Delta I$ is the difference between the photocurrent and the dark current, and *P* is the incident light power illuminated on the sample) is as high as 1.02 A/W at $V_{ds}$ = 0.5 V. The photoresponsivity deceases with the incident light power increasing (see Supplementary Section V), possibly due to the reduction of the number of carrier that could be collected under high photon flux[29]. The above results suggest the coplanar 1T′-2H-1T′ heterophase be a promising candidate for future NIR flexible and transparent optoelectronics.

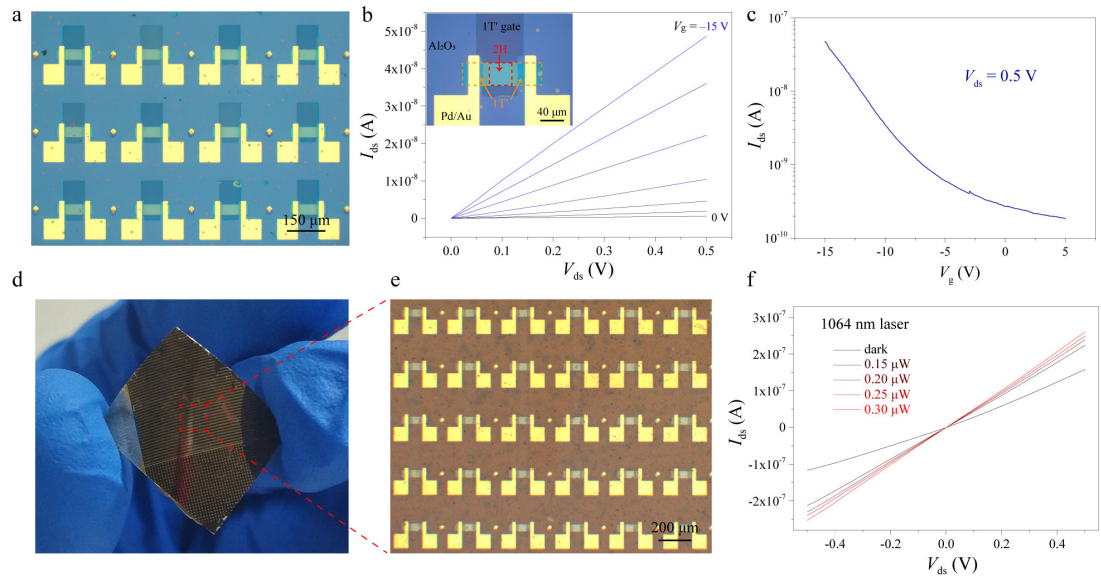

**Figure 4| 1T′ MoTe$_2$ gated heterophase transistor and the flexible NIR photodetector. a,** Optical image of the 1T′ MoTe$_2$ gated heterophase FET array. The 1T′ MoTe$_2$ local gate electrodes were fabricated by the patterned growth method. **b,** Typical $I_{ds}$-$V_{ds}$ curves of a 1T′ MoTe$_2$ gated heterophase FET measured under various gate voltages. All the $I_{ds}$-$V_{ds}$ curves exhibit linear ohmic contact behavior. Inset: zoomed-in optical image of the 1T′ MoTe$_2$ gated heterophase FET device. **c,** Transfer curve of the 1T′ MoTe$_2$ gated heterophase FET, showing a *p*-type characteristic. **d,** and **e,** Photograph and optical image of the transferred 1T′-2H-1T′ MoTe$_2$ heterophase devices on a flexible polyimide substrate. The heterostructure maintains its integrity in large scale after being transferred. **f,** The *I-V* curves of the heterophase photodetector under 1064 nm laser illumination with different incident light powers, showing obvious NIR photoresponse.

**Conclusion**

We demonstrate spatially controlled chemical assembly of integrated 1T′-2H-1T′ coplanar heterophase MoTe$_2$ FETs in a centimeter scale via phase engineered growth. Such coplanar interfaces are atomically sharp and seamless across the two phases. This approach establishes a new type of atomic-scale electrical contact. The heterophase contact FETs exhibit ohmic contact with reduced contact resistance. Besides, the field-effect mobility of the 2H MoTe$_2$ can rival that of the exfoliated single crystalline one, due to the large domain size (up to 486 μm) of the synthesized 2H MoTe$_2$. With the developed spatial-controlled patterned growth method, we fabricate the 1T′ MoTe$_2$ gated heterophase FET array, whose components of channel, gate, and contact electrodes are all 2D materials. We also transfer the heterophase device array onto a flexible substrate and demonstrate NIR photoresponse with high photoresponsivity. Our spatial-controlled large-scale chemical assembly of coplanar conductor-semiconductor heterophase structure provides a route toward industrial compatible wafer-scale high-performance 2D electronics and optoelectronics.

**Methods**

**Synthesis of the 2H and 1T′ MoTe$_2$ films**

The MoTe$_2$ films were synthesized by tellurizing the Mo films at atmospheric pressure using a horizontal tube furnace equipped with mass flow controllers and vacuum pump. Mo films were deposited on Si/SiO$_2$ substrates using magnetron sputtering. The substrates and Te powders were placed on an alumina boat, which was later inserted into a one-inch diameter quartz tube inside the furnace. After evacuating the quartz tube to less than 10 mTorr, high purity Ar gas started to flow at a rate of 500 standard cubic centimeter per minute (sccm) until atmospheric pressure was reached. After that, Ar and H$_2$ flowed at rates of 4 and 5 sccm, respectively. The furnace was ramped to 650°C for 2h and 530 °C for 30 min to synthesize 2H and 1T′ MoTe$_2$ films, respectively. After growth, the furnace was cooled down to room temperature naturally. For the patterned growth of MoTe$_2$, Mo film was pre-patterned by either photolithography or EBL followed by magnetron sputtering and lift-off process.

**Transfer of the 1T′-2H-1T′ heterophase MoTe$_2$ array**

The heterophase MoTe$_2$ array on the Si/SiO$_2$ substrate was spin-coated with a poly methyl methacrylate (PMMA) layer under 3000 rpm for 60 s. After being baked at 100 °C for 2 min, the sample was immersed in a dilute HF solution (1.5 %) at room temperature for 10 min. Subsequently, the heterophase MoTe$_2$ array with PMMA was gently peeled off from Si/SiO$_2$ substrate in deionized water and transferred onto a mesh copper grid with carbon film on top for TEM characterization or a PI substrate for flexible device investigation. Finally, the PMMA was removed using acetone, and the sample was thoroughly rinsed with isopropyl alcohol (IPA).

**Fabrication of the 1T′ MoTe$_2$ gated heterophase transistors**

Firstly, the bottom 1T′ MoTe$_2$ gate electrodes were grown by the patterned growth method on the p$^+$-Si/SiO$_2$ (285 nm) substrate. Secondly, a layer of 30 nm Al$_2$O$_3$ was deposited by atomic layer deposition (ALD). Then, a pre-grown 2H-MoTe$_2$ film was transferred onto the Al$_2$O$_3$ layer by the above described transfer method (see Method for details). Finally, the 1T′-2H-1T′ MoTe$_2$ heterophase structures were obtained by the phase engineered chemical assembly method described in the main text. A spatial-controlled photolithography is needed to define the 2H MoTe$_2$ channels on top of the 1T' MoTe$_2$ gate electrodes. For convenience of measurement, 10/50 nm Pd/Au metal electrodes were fabricated to contact the 1T′ MoTe$_2$.

## Author Contribution

Y.Y. conceived the project. Y.Y., L.D. and X.L.X. designed the experiment. X.L.X. fabricated the devices and performed electrical measurements. S.L. and X.L.X. synthesized the material. B.H. and P.G. conducted the STEM characterizations. Y.M.H. and D.M. conducted the HR-TEM and SAED characterizations. W.J.X. did the magnetron sputtering. X.H.Y. helped to perform the photodetection measurement. K.Y. conducted the E-beam evaporation. P.L., W.T.G. and S.Q.Y. contributed to the schematic illustrations. Y.Y. and L.D. supervised this research. X.L.X., Y.Y. and L.D. wrote the manuscript. All authors contributed to discussions.

## Acknowledgement

This work was supported by the National Natural Science Foundation of China (Nos. 61874003, 61521004, 11474007), Nation Key R&D Program of China (Grant Nos. 2018YFA0306900 and 2017YFA0206301), Beijing Natural Science Foundation (4182028), and the "1000 Youth Talent Plan" Fund.

## References

1. Lin, Z.; Liu, Y.; Halim, U.; Ding, M.; Liu, Y.; Wang, Y.; Jia, C.; Chen, P.; Duan, X.; Wang, C.; Song, F.; Li, M.; Wan, C.; Huang, Y.; Duan, X. Solution-Processable 2d Semiconductors for High-Performance Large-Area Electronics. *Nature* **2018,** *562*, 254-258.
2. Kang, K.; Lee, K.-H.; Han, Y.; Gao, H.; Xie, S.; Muller, D. A.; Park, J. Layer-by-Layer Assembly of Two-Dimensional Materials into Wafer-Scale Heterostructures. *Nature* **2017,** *550*, 229.
3. Yu, H.; Liao, M.; Zhao, W.; Liu, G.; Zhou, X. J.; Wei, Z.; Xu, X.; Liu, K.; Hu, Z.;

Deng, K.; Zhou, S.; Shi, J.-A.; Gu, L.; Shen, C.; Zhang, T.; Du, L.; Xie, L.; Zhu, J.; Chen, W.; Yang, R.; Shi, D.; Zhang, G. Wafer-Scale Growth and Transfer of Highly-Oriented Monolayer Mos2 Continuous Films. *ACS Nano* **2017,** *11*, 12001-12007.

4. Xue, Y.; Zhang, Y.; Liu, Y.; Liu, H.; Song, J.; Sophia, J.; Liu, J.; Xu, Z.; Xu, Q.; Wang, Z.; Zheng, J.; Liu, Y.; Li, S.; Bao, Q. Scalable Production of a Few-Layer Mos2/Ws2 Vertical Heterojunction Array and Its Application for Photodetectors. *ACS Nano* **2016,** *10*, 573-580.

5. Li, L.; Yang, F.; Ye, G. J.; Zhang, Z.; Zhu, Z.; Lou, W.; Zhou, X.; Li, L.; Watanabe, K.; Taniguchi, T.; Chang, K.; Wang, Y.; Chen, X. H.; Zhang, Y. Quantum Hall Effect in Black Phosphorus Two-Dimensional Electron System. *Nature Nanotechnology* **2016,** *11*, 593.

6. Bandurin, D. A.; Tyurnina, A. V.; Yu, G. L.; Mishchenko, A.; Zólyomi, V.; Morozov, S. V.; Kumar, R. K.; Gorbachev, R. V.; Kudrynskyi, Z. R.; Pezzini, S.; Kovalyuk, Z. D.; Zeitler, U.; Novoselov, K. S.; Patanè, A.; Eaves, L.; Grigorieva, I. V.; Fal'ko, V. I.; Geim, A. K.; Cao, Y. High Electron Mobility, Quantum Hall Effect and Anomalous Optical Response in Atomically Thin Inse. *Nature Nanotechnology* **2016,** *12*, 223.

7. Qiu, G.; Wang, Y.; Nie, Y.; Zheng, Y.; Cho, K.; Wu, W.; Ye, P. D. Quantum Transport and Band Structure Evolution under High Magnetic Field in Few-Layer Tellurene. *Nano Letters* **2018,** *18*, 5760-5767.

8. Allain, A.; Kang, J.; Banerjee, K.; Kis, A. Electrical Contacts to Two-Dimensional Semiconductors. *Nature Materials* **2015,** *14*, 1195.

9. English, C. D.; Shine, G.; Dorgan, V. E.; Saraswat, K. C.; Pop, E. Improved Contacts to Mos2 Transistors by Ultra-High Vacuum Metal Deposition. *Nano Letters* **2016,** *16*, 3824-3830.

10. Liu, Y.; Guo, J.; Zhu, E.; Liao, L.; Lee, S.-J.; Ding, M.; Shakir, I.; Gambin, V.; Huang, Y.; Duan, X. Approaching the Schottky–Mott Limit in Van Der Waals Metal–Semiconductor Junctions. *Nature* **2018,** *557*, 696-700.

11. Kappera, R.; Voiry, D.; Yalcin, S. E.; Branch, B.; Gupta, G.; Mohite, A. D.; Chhowalla, M. Phase-Engineered Low-Resistance Contacts for Ultrathin Mos2 Transistors. *Nature Materials* **2014,** *13*, 1128.

12. Cho, S.; Kim, S.; Kim, J. H.; Zhao, J.; Seok, J.; Keum, D. H.; Baik, J.; Choe, D.-H.; Chang, K. J.; Suenaga, K.; Kim, S. W.; Lee, Y. H.; Yang, H. Phase Patterning for Ohmic Homojunction Contact in MoTe$_2$. *Science* **2015,** *349*, 625.

13. Sung, J. H.; Heo, H.; Si, S.; Kim, Y. H.; Noh, H. R.; Song, K.; Kim, J.; Lee, C. S.; Seo, S. Y.; Kim, D. H.; Kim, H. K.; Yeom, H. W.; Kim, T. H.; Choi, S. Y.; Kim, J. S.; Jo, M. H. Coplanar Semiconductor-Metal Circuitry Defined on Few-Layer MoTe$_2$ Via Polymorphic Heteroepitaxy. *Nat Nanotechnol* **2017,** *12*, 1064-1070.

14. Leong, W. S.; Ji, Q.; Mao, N.; Han, Y.; Wang, H.; Goodman, A. J.; Vignon, A.; Su, C.; Guo, Y.; Shen, P.-C.; Gao, Z.; Muller, D. A.; Tisdale, W. A.; Kong, J. Synthetic Lateral Metal-Semiconductor Heterostructures of Transition Metal Disulfides. *Journal of the American Chemical Society* **2018,** *140*, 12354-12358.

15. Octon, T. J.; Nagareddy, V. K.; Russo, S.; Craciun, M. F.; Wright, C. D. Fast High-Responsivity Few-Layer MoTe2 Photodetectors. *Advanced Optical Materials* **2016,** *4*, 1750-1754.

16. Zhang, K.; Fang, X.; Wang, Y.; Wan, Y.; Song, Q.; Zhai, W.; Li, Y.; Ran, G.; Ye, Y.; Dai, L. Ultrasensitive near-Infrared Photodetectors Based on a Graphene–MoTe2–Graphene Vertical Van Der Waals Heterostructure. *ACS Applied Materials & Interfaces* **2017,** *9*, 5392-5398.

17. Bie, Y.-Q.; Grosso, G.; Heuck, M.; Furchi, M. M.; Cao, Y.; Zheng, J.; Bunandar, D.; Navarro-Moratalla, E.; Zhou, L.; Efetov, D. K.; Taniguchi, T.; Watanabe, K.; Kong, J.; Englund, D.; Jarillo-Herrero, P. A MoTe2-Based Light-Emitting Diode and Photodetector for Silicon Photonic Integrated Circuits. *Nature Nanotechnology* **2017,** *12*, 1124.

18. Duerloo, K.-A. N.; Li, Y.; Reed, E. J. Structural Phase Transitions in Two-Dimensional Mo- and W-Dichalcogenide Monolayers. *Nature Communications* **2014,** *5*, 4214.

19. Li, Y.; Duerloo, K.-A. N.; Wauson, K.; Reed, E. J. Structural Semiconductor-to-Semimetal Phase Transition in Two-Dimensional Materials Induced by Electrostatic Gating. *Nature Communications* **2016,** *7*, 10671.

20. Song, S.; Keum, D. H.; Cho, S.; Perello, D.; Kim, Y.; Lee, Y. H. Room Temperature Semiconductor–Metal Transition of MoTe2 Thin Films Engineered by Strain. *Nano Letters* **2016,** *16*, 188-193.

21. Keum, D. H.; Cho, S.; Kim, J. H.; Choe, D.-H.; Sung, H.-J.; Kan, M.; Kang, H.; Hwang, J.-Y.; Kim, S. W.; Yang, H.; Chang, K. J.; Lee, Y. H. Bandgap Opening in Few-Layered Monoclinic MoTe2. *Nature Physics* **2015,** *11*, 482.

22. Xu, X.; Chen, S.; Liu, S.; Cheng, X.; Xu, W.; Li, P.; Wan, Y.; Yang, S.; Gong, W.; Yuan, K.; Gao, P.; Ye, Y.; Dai, L. Millimeter-Scale Single-Crystalline Semiconducting Mote2 Via Solid-to-Solid Phase Transformation. *Journal of the American Chemical Society* **2019,** *141*, 2128-2134.

23. Blake, P.; Hill, E. W.; Castro Neto, A. H.; Novoselov, K. S.; Jiang, D.; Yang, R.; Booth, T. J.; Geim, A. K. Making Graphene Visible. *Applied Physics Letters* **2007,** *91*, 063124.

24. Zhou, L.; Xu, K.; Zubair, A.; Liao, A. D.; Fang, W.; Ouyang, F.; Lee, Y.-H.; Ueno, K.; Saito, R.; Palacios, T.; Kong, J.; Dresselhaus, M. S. Large-Area Synthesis of High-Quality Uniform Few-Layer Mote2. *Journal of the American Chemical Society* **2015,** *137*, 11892-11895.

25. Yun, S. J.; Han, G. H.; Kim, H.; Duong, D. L.; Shin, B. G.; Zhao, J.; Vu, Q. A.; Lee, J.; Lee, S. M.; Lee, Y. H. Telluriding Monolayer Mos2 and Ws2 Via Alkali Metal Scooter. *Nature Communications* **2017,** *8*, 2163.

26. Tang, H.-L.; Chiu, M.-H.; Tseng, C.-C.; Yang, S.-H.; Hou, K.-J.; Wei, S.-Y.; Huang, J.-K.; Lin, Y.-F.; Lien, C.-H.; Li, L.-J. Multilayer Graphene–Wse2 Heterostructures for Wse2 Transistors. *ACS Nano* **2017,** *11*, 12817-12823.

27. Lee, J. S.; Choi, S. H.; Yun, S. J.; Kim, Y. I.; Boandoh, S.; Park, J.-H.; Shin, B. G.; Ko, H.; Lee, S. H.; Kim, Y.-M.; Lee, Y. H.; Kim, K. K.; Kim, S. M. Wafer-Scale Single-Crystal Hexagonal Boron Nitride Film Via Self-Collimated Grain Formation. *Science* **2018,** *362*, 817.

28. Roy, T.; Tosun, M.; Kang, J. S.; Sachid, A. B.; Desai, S. B.; Hettick, M.; Hu, C. C.; Javey, A. Field-Effect Transistors Built from All Two-Dimensional Material Components. *ACS Nano* **2014,** *8*, 6259-6264.

29. Buscema, M.; Groenendijk, D. J.; Blanter, S. I.; Steele, G. A.; van der Zant, H. S. J.; Castellanos-Gomez, A. Fast and Broadband Photoresponse of Few-Layer Black Phosphorus Field-Effect Transistors. *Nano Letters* **2014,** *14*, 3347-3352.